\documentclass[aip,showpacs]{revtex4-1}
\usepackage{graphicx}
\usepackage{color}

\usepackage[version=3]{mhchem} 
\usepackage{gensymb}
\usepackage{amssymb}
\usepackage{upgreek}
\usepackage[usenames,dvipsnames]{xcolor}
\usepackage{natbib}

\begin{document}

\title [] {Quantum Emission From Hexagonal Boron Nitride Monolayers}

\author{Toan Trong Tran}
\affiliation{School of Physics and Advanced Materials, University of Technology, Sydney, 15 Broadway, Ultimo, New South Wales 2007, Australia}

\author{Kerem Bray}
\affiliation{School of Physics and Advanced Materials, University of Technology, Sydney, 15 Broadway, Ultimo, New South Wales 2007, Australia}

\author{Michael J. Ford}
\affiliation{School of Physics and Advanced Materials, University of Technology, Sydney, 15 Broadway, Ultimo, New South Wales 2007, Australia}

\author{Milos Toth}
\email{Milos.Toth@uts.edu.au}
\affiliation{School of Physics and Advanced Materials, University of Technology, Sydney, 15 Broadway, Ultimo, New South Wales 2007, Australia}

\author{Igor Aharonovich}
\email{Igor.Aharonovich@uts.edu.au}
\affiliation{School of Physics and Advanced Materials, University of Technology, Sydney, 15 Broadway, Ultimo, New South Wales 2007, Australia}

\date{\today}

\begin{abstract}
Atomically thin van der Waals crystals have recently enabled new scientific and technological breakthroughs across a variety of disciplines in materials science, nanophotonics and physics. However, non-classical photon emission from these materials has not been achieved to date. Here we report room temperature quantum emission from hexagonal boron nitride nanoflakes. The single photon emitter exhibits a combination of superb quantum optical properties at room temperature that include the highest brightness reported in the visible part of the spectrum, narrow line width, absolute photo-stability, a short excited state lifetime and a high quantum efficiency. Density functional theory modeling suggests that the emitter is the antisite nitrogen vacancy defect that is present in  single and multi-layer hexagonal boron nitride. Our results constitute the unprecedented potential of van der Waals crystals for nanophotonics, optoelectronics and quantum information processing.
\end{abstract}

\maketitle

Recent progress in the exploration of van der Walls crystals\cite{Geim2013} has resulted in successful synthesis of a new generation of two dimensional (2D) materials including molybdenum disulphide\cite{Mak2010,Coleman2011}, tungsten disulphide\cite{Matte2010,Zeng2011}, hexagonal boron nitride (hBN)\cite{Shi2010,Song2010,Ismach2012,Kim2012a,Park2014} and others\cite{Coleman2011,Novoselov2005,Miro2014}. These new atomically thin layers enabled realization of lasing from single monolayers\cite{Wu2015}, generation of phonon polaritons\cite{Dai2014a,Xu2014a,Caldwell2014}, super resolution imaging and spin valley transport\cite{Mak2012,Zeng2012,Hu2014a}. Yet, an aspiring goal is to employ these materials in the quantum regime, where the photon emission can be non classical, thereby enabling a major paradigm shift towards applications in quantum information processing\cite{Awschalom2013}.

In this work we identify and study single photon emission from localized defects in monolayer hBN and few-layer hBN flakes. The emitters are optically active at room temperature, thereby heralding the transformational application of 2D materials in quantum information applications with the potential to achieve scalable nanophotonic circuits on a single chip. 

\begin{figure}[!t]
\centering
\includegraphics[width=0.9\textwidth]{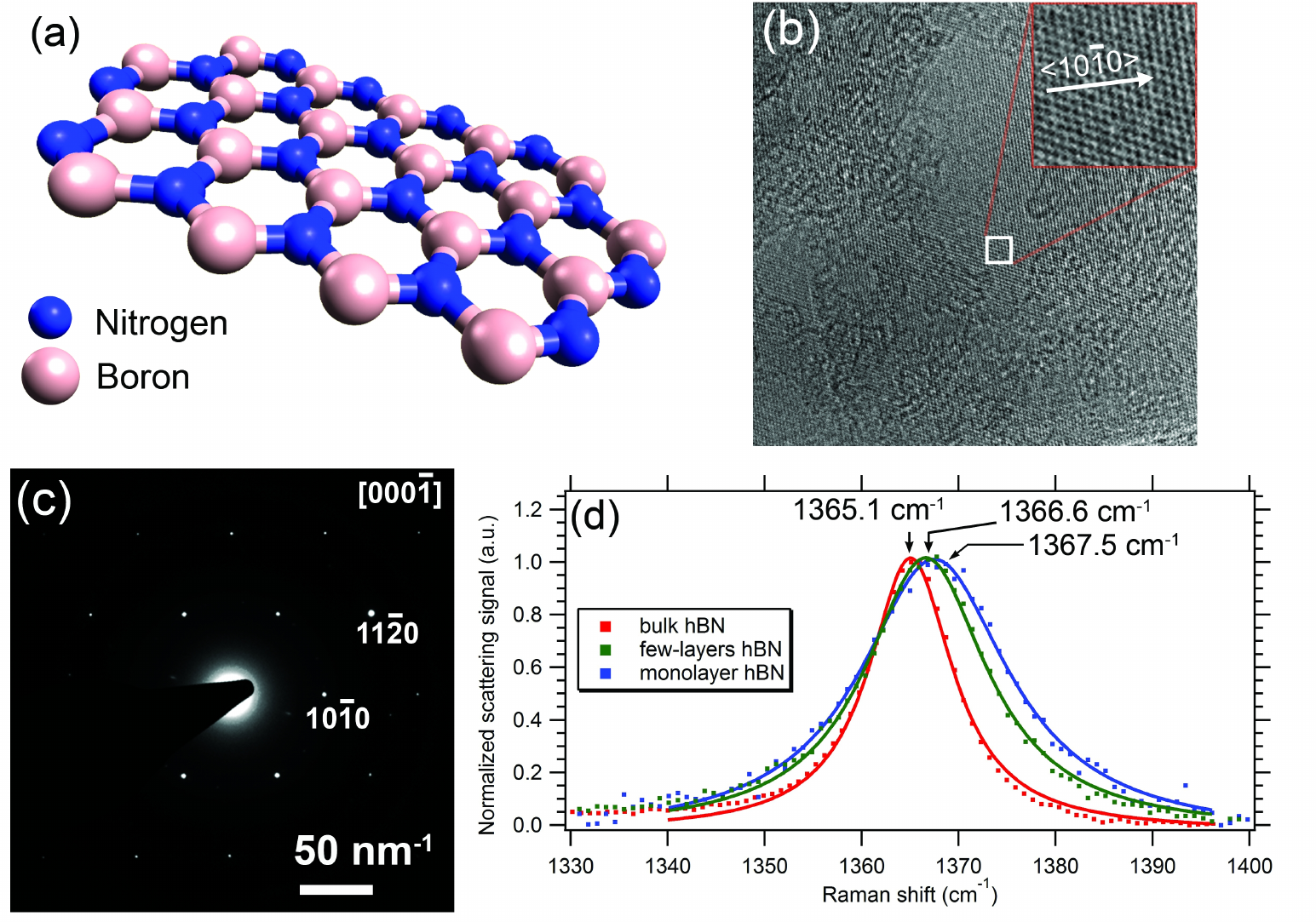}
\caption{\textbf{Structural characterization of hBN.} (a) Schematic illustration of a hBN monolayer. (b) TEM image of the corner of a single hBN sheet. The inset shows the hBN lattice. (c) a corresponding SAED pattern. (d) Raman scattering spectra of monolayer, few-layer and bulk hBN (blue, green and red squares, respectively) on a silicon substrate. Solid lines are Lorentzian fits to the experimental data.}
\label{figure1}
	\end{figure}

Being a van der Walls crystal, monolayer hBN posses similar structure to graphene where in plane bonds are much stronger than out of plane bonds\cite{Xu2013a,Geim2013}. Figure $\ref{figure1}$a shows the 2D hexagonal lattice of hBN comprised of boron and nitrogen atoms. Figure $\ref{figure1}$b and figure $\ref{figure1}$c show a transmission electron microscope image of hBN and a corresponding selected area electron diffraction (SAED) pattern. The SAED pattern exhibits a sixfold symmetry attributed to single crystal hexagonal boron nitride with a [000$\bar{1}$] viewing zone axis \cite{Song2010,Shi2010,Khan2015}. Figure $\ref{figure1}$d shows Raman spectra of a hBN monolayer, few-layer hBN and a bulk hBN reference sample. The Raman shifts are $1367.5\:cm^{-1}$ (FWHM of 18.2$\:cm^{-1}$), $1366.6\:cm^{-1}$, (FWHM of 15.2$\:cm^{-1}$) and $1365.1\:cm^{-1}$ (FWHM of 10.3$\:cm^{-1}$), respectively. The difference between the position and width of the monolayer and bulk hBN peaks are in excellent agreement with a prior study\cite{Gorbachev2011}, confirming that the single sheets used here are indeed atomically thin.

\begin{figure}[!t]
\centering
\includegraphics[width=0.9\textwidth]{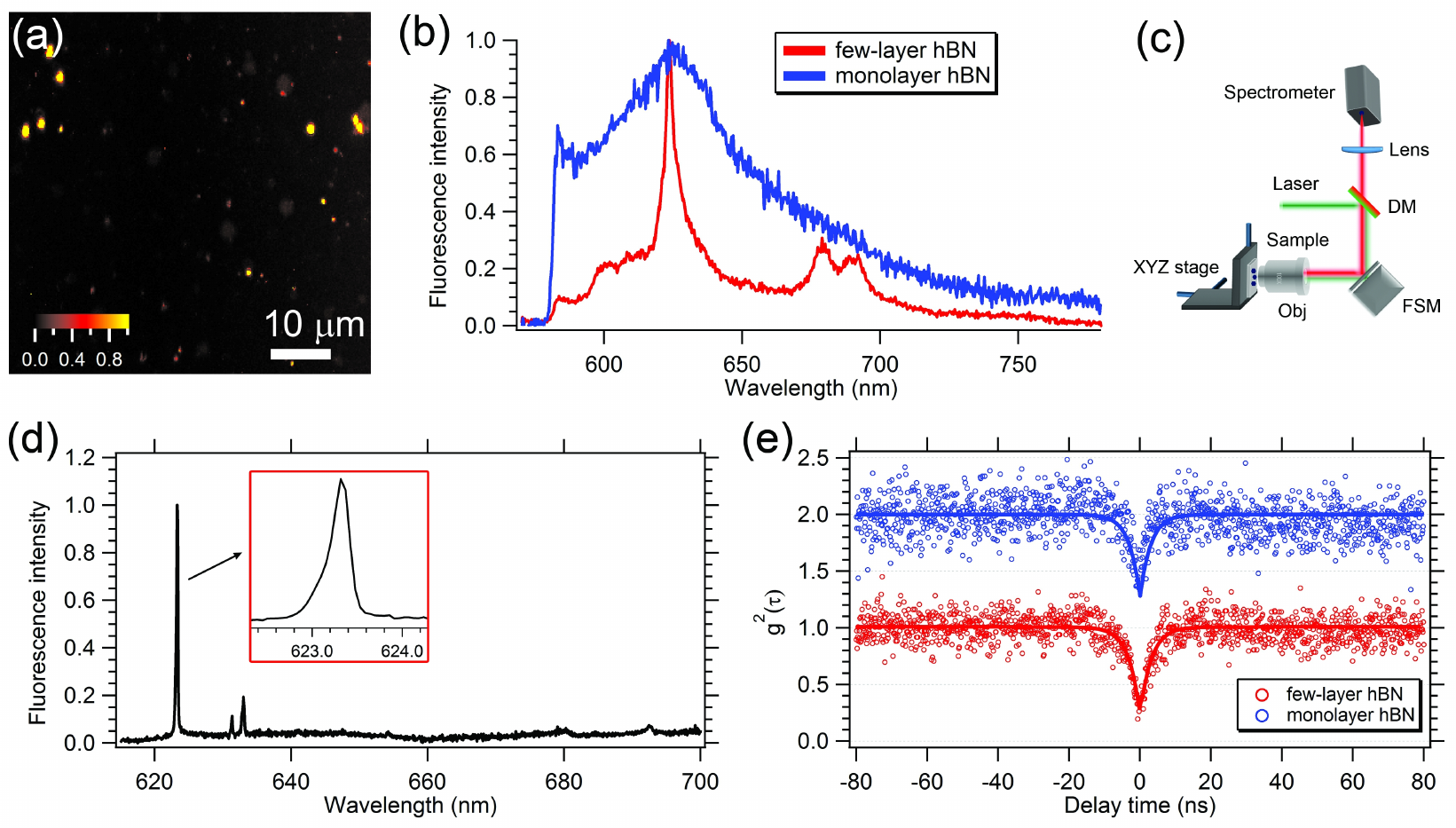}
\caption{\textbf{Optical characterization of single photon emitters in hBN.}  (a) Scanning confocal map of a few-layer hBN sample showing bright luminescent spots that correspond to emission from single defects. (b) Room temperature PL spectra of a defect center in a hBN monolayer (blue trace) and a few-layer hBN flake (red trace). (c) Schematic illustration of the confocal PL setup (DM - dichroic mirror, FSM - fast steering mirror, Obj-objective lens). (d) PL spectra taken at 77K of a defect center in a few-layer hBN flake. The inset shows the zero-phonon line. (E) Typical antibunching curves acquired from an individual defect center in a hBN monolayer (blue, open circles, shifted for clarity) and a few-layer hBN flake (red, open circles). The solid blue and red lines are fits obtained using equation $\ref{equation1}$.}
\label{figure2} 
\end{figure}

Hexagonal boron nitride has a wide bandgap of nearly 6 eV\cite{Xia2014}, which is much greater than that of most 2D materials, and is expected to host optically active defects that have  ground and excited states within the gap. To address such defects, optical measurements were performed using a low energy excitation laser emitting at 532 nm. The use of a sub-bandgap excitation source is key to probing individual defects whilst avoiding excitonic emissions\cite{Watanabe2004}. Figure $\ref{figure2}$a shows a confocal photoluminescence (PL) map recorded at room temperature from hBN flakes dispersed on a silicon dioxide substrate (see methods).  The majority of the bright spots seen in the map are attributed to single defect centers hosted by hBN flakes. Figure $\ref{figure2}$b shows PL spectra recorded from a hBN monolayer and a few-layer hBN flake. In both cases, a bright emission is detected at $\sim$623 nm (1.99 eV), attributed to the zero phonon line (ZPL) of the defect. The emission lines of few layer hBN are much narrower than those of monolayers, likely due to reduced strain (consistent with the Raman spectra shown in figure $\ref{figure1}$d). Spectra from few-layer hBN exhibit a phonon side band (PSB) doublet with well-resolved peaks at 680 nm (1.83 eV) and 693 nm (1.79 eV), respectively. Figure $\ref{figure2}$c shows a simplified confocal microscope configuration used for the current measurements. Figure $\ref{figure2}$d shows the low temperature spectrum of the defect. The ZPL narrows dramatically, and has a full width at half maximum (FWHM) of 0.25 nm. 

To demonstrate single photon emission using the same luminescence signals, we recorded second order autocorrelation functions ($g^2(\tau)$) using a Hanbury Brown and Twiss (HBT) interferometry setup. For a true single photon source, the  $g^2(\tau)$ curve dips below 0.5 at zero delay time ($\tau$ = 0)\cite{Kurtsiefer2000,Michler2000a,Lounis2005}. Representative antibunching curves of the defect center in  monolayer hBN (blue, open circles) and few-layer hBN (red, open circles) are shown in figure $\ref{figure2}$e. In both cases, a dip below 0.5 is clearly seen at zero delay time ($\tau$ = 0), confirming that each defect acts as a single photon emitter\cite{Kurtsiefer2000,Michler2000a,Lounis2005}. The results serve as direct evidence for quantum emission from a single defect in hBN monolayers and in few layer hBN. The experimental data were fit using a three-level model (equation $\ref{equation1}$). The parameters $\tau_1$ and $\tau_2$ are the lifetimes of the excited and metastable states, respectively. The fluorescence lifetime of the defect from these measurements is $\sim$ 2.5 ns. A detailed analysis of this quantum system is presented in the supporting information.   

\begin{equation}
g^2(\tau) = 1 - (1+a)e^{-{\vert\tau\vert}/{\tau_1}} + ae^{-{\vert\tau\vert}/{\tau_2}}
\label{equation1}
\end{equation}

To understand the nature of the defect, detailed photophysical characterization of the defects in few-layer hBN was performed. First, the Debye-Waller (DW) factor was calculated to understand the extent of electron-phonon coupling, i.e. the ratio of the ZPL intensity to that of the total emission \cite{Elke2011}. Based on the PL measurements presented in figure $\ref{figure2}$, the DW factor is $\sim0.82$. This value is amongst the highest reported for quantum emitters\cite{Aharonovich2014}. Centers with high DW factors are very useful for many applications including optoelectronics, nanophotonics and bio-imaging, where there is a need for a strong, narrow band signal and a high signal to noise ratio.

\begin{figure}[!t]
\centering
\includegraphics[width=0.9\textwidth]{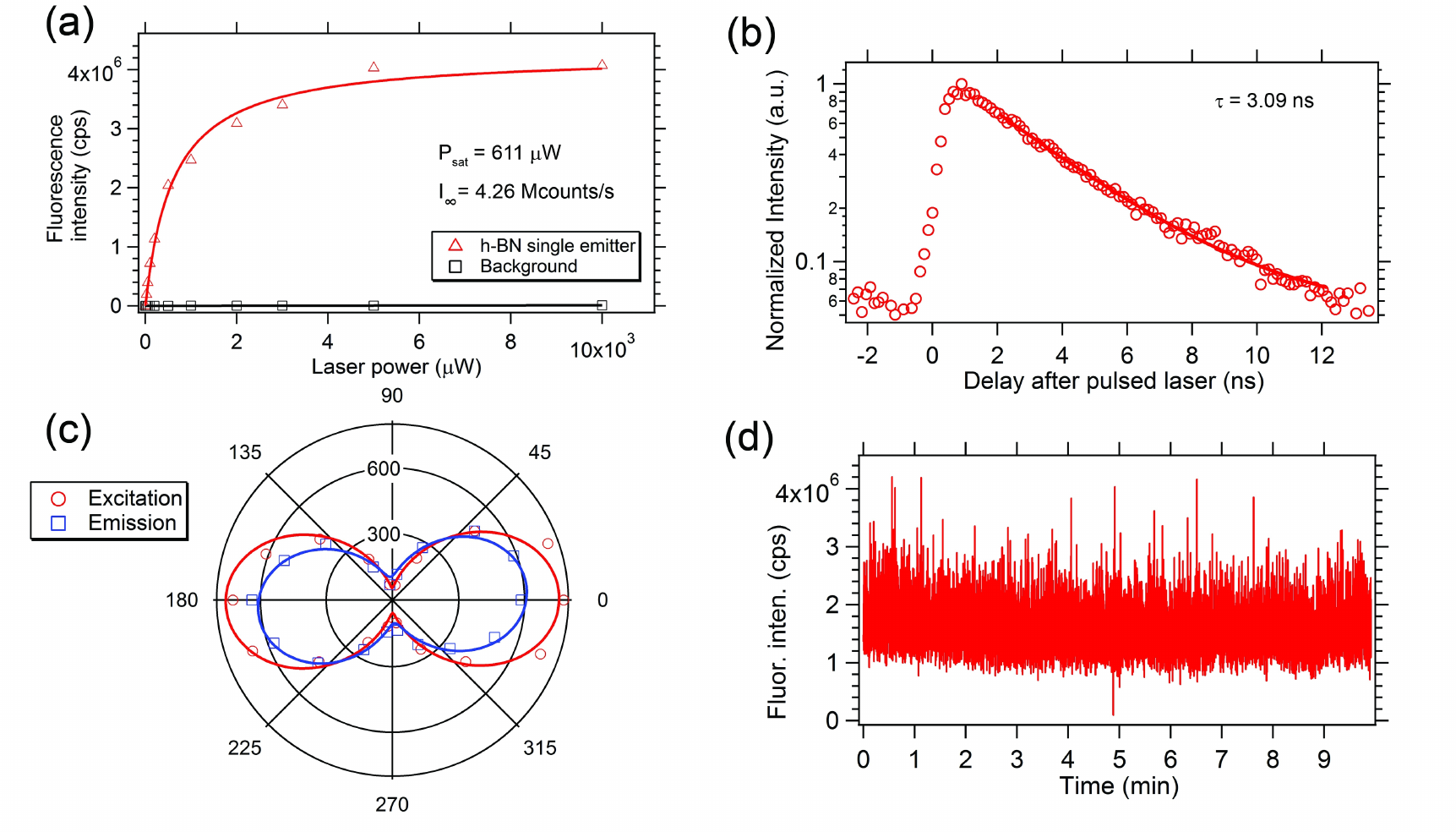}
\caption{\textbf{Photophysical properties of single photon emitters in few-layer hBN.} (a) Fluorescence saturation curve obtained from a single defect, showing a maximum emission rate of 4.26 Mcps. (b) Fluorescence lifetime measurement exhibiting an excited state lifetime of 3.1 ns. (c) Excitation (open, red circles) and emission (open, blue squares) polarization curves from a single defect. The solid red and blue lines are fits obtained using a $cos^{2}$($\theta$) function. (d) Fluorescence intensity as a function of time demonstrating the photostability of a single defect. All the measurments are performed at room temperature using a 532 nm excitation laser.}
\label{figure3}
	\end{figure}

Figure $\ref{figure3}$a shows the emission intensity of a single emitter as a function of excitation power. The experimental data were fit using \textit {I=I$_\infty$/(P+P$_{sat}$)}, where $I_\infty$, P$_{sat}$ are the  emission rate and excitation power at saturation, respectively. The fitting yields $I_\infty \sim$  4.2 x 10$^{6}$ counts/s at P$_{sat} \sim$ 611 $\mu$W. This brightness is the highest reported for a quantum emitter in the visible spectral range, and similar to that of the brightest quantum emitters known to date\cite{Aharonovich2009,Elke2011}. While several other systems exhibit comparable count rates, the availability of hBN material in bulk quantities will enable wide usage of this material in emerging technologies 
Time-resolved fluorescence measurements were performed to directly measure the excited state lifetime. Figure $\ref{figure3}$b shows the decay of single quantum emitter with a ZPL at 634 nm. The experimental data fits perfectly with a single exponential, and yields a lifetime of $\sim3$ ns, in accord with the lifetime values deduced by measuring the second order correlation function.

To understand whether the emission center consists of a single or multiple dipoles, we performed excitation and emission polarization measurements. Figure $\ref{figure3}$c shows the excitation (red circles) and emission (blue squares) polarization data obtained from a typical single emitter, and the corresponding fits obtained using a $cos^{2}$($\theta$) fitting function\cite{Schietinger2009}. The excitation and emission polarization visibility are calculated to be 86$\%$ and 71$\%$, respectively. These polarization results are characteristic of a single linearly polarized dipole transition.

The stability of a typical single emitter is shown in figure $\ref{figure3}$d. Even under a high excitation power of 3 mW, the defect exhibits a stable fluorescence emission signal over more than 10 min without any blinking or bleaching. Interestingly, while the emitters embedded in few-layer hBN exhibit absolute photo-stability, the emitters in monolayer hBN blink and bleach after several excitation cycles. We tentatively ascribe this behavior to modification of the defects caused by direct contact with the environment.

To complete our characterization of the photophysical properties, we calculate the quantum efficiency for the quantum emitter. Combining the analysis of the saturation, polarization and the auto-correlation data, the quantum efficiency is approximated at $\sim$65$\%$. This value, along with a short radiative lifetime of $\sim$3ns implies that the center is one the most efficient single photon emitters reported to date\cite{Neu2012a}\cite{Aharonovich2010}. The details of the calculations and the extended three-level model are presented in the Supplementary Information. 

We now turn to the origin of the defect responsible for the observed single photon emission. Given that the quantum emission from this defect is observed in monolayers and few-layer hBN materials, the defect is most likely intrinsic. The most likely candidates are a nitrogen vacancy (V$_{N}$), a boron vacancy (V$_{B}$) or an anti-site complex in which the nitrogen occupies the boron site and there is a missing atom at the nitrogen cite (N$_{B}$V$_{N}$)\cite{Orellana2001,Attaccalite2011}. We exclude the possibility of a di-vacancy as those have been shown to be highly unstable\cite{Jin2009}. The V$_{B}$ has been theoretically predicted to have a UV absorption and emission band\cite{Attaccalite2011} that is inconsistent with our experimental data. We therefore used density function theory (DFT) to investigate the V$_{N}$ and the N$_{B}$V$_{N}$ defects using the Vienna ab initio simulation package VASP \cite{Kresse1994,Kresse1996,Kresse1996a} with the Perdew-Burke-Ernzerhof (PBE) exchange-correlation functional\cite{Perdew1996}.

\begin{figure}[!t]
\centering
\includegraphics[width=0.9\textwidth]{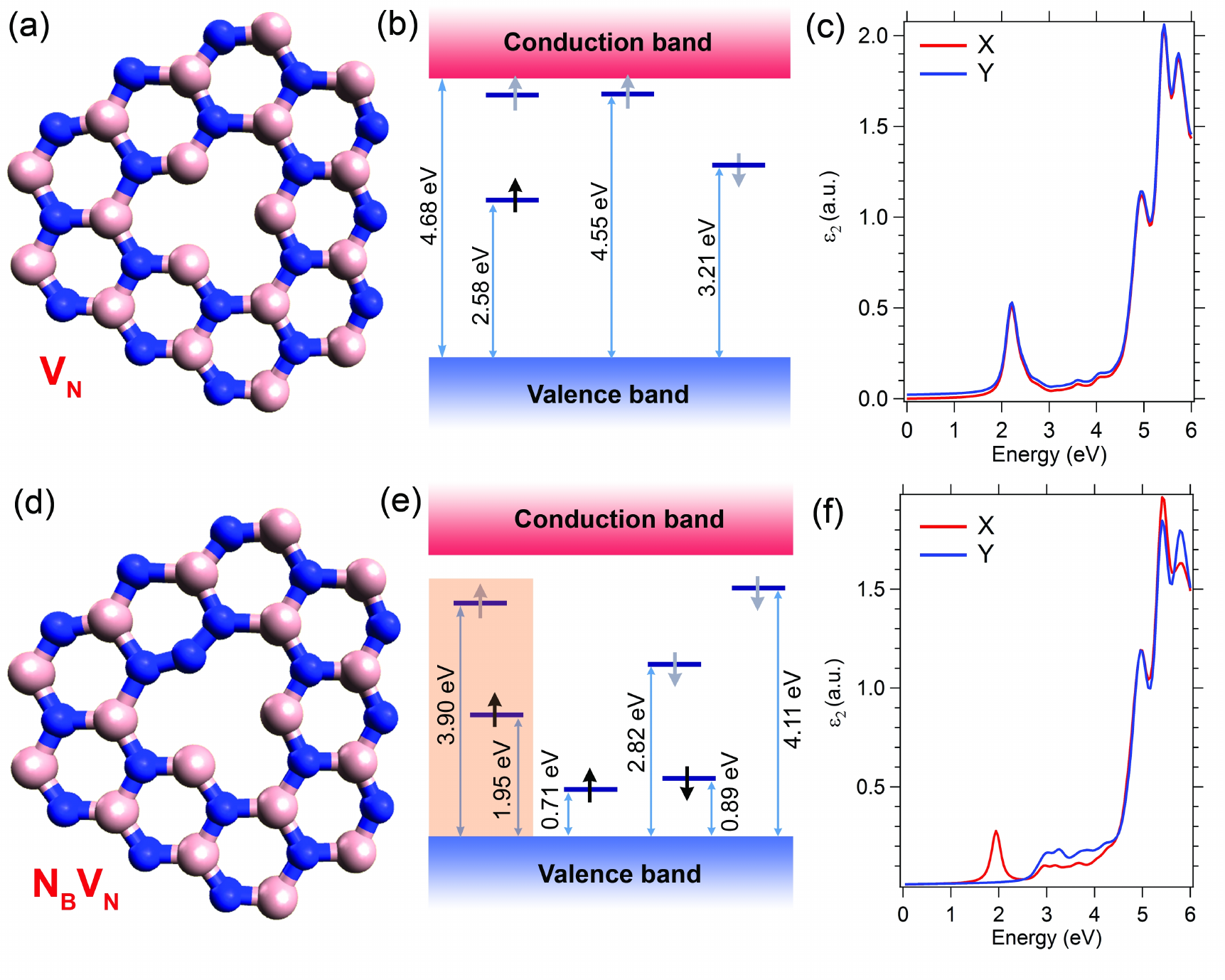}
	\caption{\textbf{Proposed defect models in hBN lattice.} Schematics of (a) Nitrogen-vacancy, V$_{N}$ and (d) anti-site nitrogen vacancy, N$_B$V$_N$. (b) Simulated electronic structures of the nitrogen-vacancy, V$_{N}$, and (e) anti-site nitrogen vacancy, N$_B$V$_N$. Black and grey arrows represent occupied and unoccupied states, respectively. (c) Calculated imaginary dielectric tensors of the two components, namely, X (red trace) and Y (blue trace) of  nitrogen-vacancy, V$_{N}$ and (f) anti-site nitrogen vacancy, N$_B$V$_N$.}
\label{figure4}	
\end{figure}

Figure $\ref{figure4}$a,c show the structure of the V$_{N}$ and N$_{B}$V$_{N}$ defects, respectively. The V$_{N}$ center has a missing nitrogen atom and the N$_{B}$V$_{N}$ defect has a nitrogen atom at a Boron cite neighboring a vacancy.

Figure $\ref{figure4}$b,d present the corresponding energy levels of the V$_{N}$ and N$_{B}$V$_{N}$ (figure $\ref{figure4}$d) centers.  Bearing in mind the shortcomings of DFT in predicting bandgaps and that the PBE is expected to underestimate this quantity, these calculations predict a number of transitions that are close to the measured photon energy.  We assume that only spin preserving transitions are allowed. For the V$_{N}$ defect there is a single transition between a potential ground state 2.58 eV above the valence band to a potential excited state located at 4.55 eV above the valence band, resulting in a 1.97 eV transition. The N$_{B}$V$_{N}$ defect exhibits potential internal transitions for the two spin channels that are nearly degenerate at 1.95 eV and 1.93 eV (shaded in figure $\ref{figure4}$e) . Both of the two proposed centers, therefore, possess theoretical transition energies that match our experimental data (a ZPL centered on 630 nm, corresponding to $\sim$ 2 eV).

To distinguish between the two defect possibilities, we simulate the optical responses of the defects. Figure $\ref{figure4}$c,f  show the calculated imaginary dielectric tensor for two components in the plane of the 2D sheet.  The dielectric function corresponding to the V$_{N}$ defect is isotropic and shows a single peak within the bandgap at 2.1 eV, consistent with a transition from the occupied defect level into the conduction band. The internal ground-to-excited state transition is either forbidden or extremely weak and hidden in the low-energy side of this transition. Therefore, the V$_{N}$ defect is unlikely to be the studied emitter. 

The N$_{B}$V$_{N}$ defect is, by contrast, highly anisotropic and exhibits a peak at approximately 1.9 eV, consistent with the internal ground-to-excited state transitions. This theoretical prediction is consistent with our experimental data, suggesting that the N$_{B}$V$_{N}$ defect is the most likely candidate for the quantum emitter within hBN monolayers. Moreover, a spin resolved calculation shows that the N$_{B}$V$_{N}$ energy levels responsible for the optical transition are the ones between the 1.95 eV (ground state) and 3.90 eV (excited state) above the valence band maximum. This transition is highlighted in figure $\ref{figure4}$e. The spin-resolved dielectric functions are provided in the supporting information.

In order to ascertain that the simulation is also valid for few-layer hBN, we performed the same calculations using three-layer hBN where the middle layer contains the defect. As expected, the results show that the weak interaction between the sheets has little effect, yielding an energy level structure that is  very similar to that plotted in figure $\ref{figure4}$ (see supplementary information for details). 

To summarize, we unveiled the quantum emission from a point defect, N$_{B}$V$_{N}$, in monolayer hBN and few layer hBN. The emitters are fully polarized and exhibit record rates of $\sim$ 4 million counts/s at room temperature. Our results augment the scientific and technological importance of Van der Walls crystals - in particular hBN. Hosting single emitters within hBN will enable new applications in quantum technologies and optoelectornics employing 2D materials and highlight the emerging potential of hBN materials.\\

{\bf{Acknowledgments}}

We thanks Shay Lifshitz for useful discussions and Jinghua Fang for assistance with TEM images. The work was supported in part by the Australian Research Council (Project Number DP140102721) and FEI Company. I. A. is the recipient of an Australian Research Council Discovery Early Career Research Award (Project Number DE130100592). 

\bibliography{Photonics}

\end{document}